\documentclass[fleqn,twoside]{article}
\usepackage{espcrc2,epsfig}

\usepackage{graphicx}
\usepackage[figuresright]{rotating}

\newcommand{\AmS}{{\protect\the\textfont2
  A\kern-.1667em\lower.5ex\hbox{M}\kern-.125emS}}
\newcommand{\agt}{\matrix{>\cr\noalign{\vskip-7pt}\sim\cr}}
\newcommand{\alt}{\matrix{<\cr\noalign{\vskip-7pt}\sim\cr}}
\newcommand{\notE}{\ \hbox{{$E$}\kern-.60em\hbox{/}}}
\newcommand{\notp}{\ \hbox{{$p$}\kern-.43em\hbox{/}}}
\title{
\begin{flushright}
\includegraphics{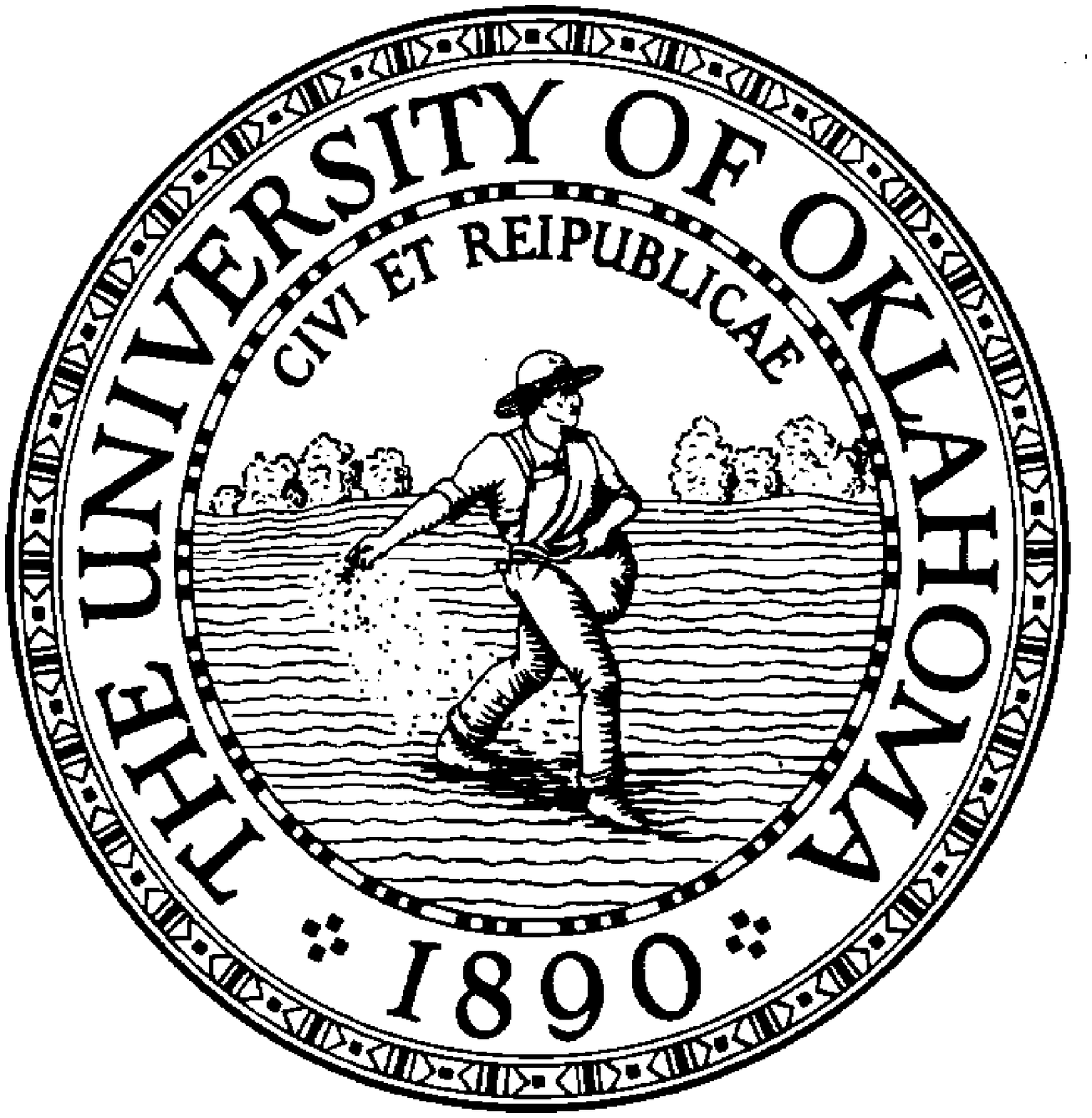}
{\small\bf OKHEP-02-06} \\
\hspace{2in} The University of Oklahoma \hfill {\small\bf hep-ph/0210001} \\
{\small September 2002} 
\end{flushright}
Higgs Decays into Leptons\thanks{
Presented at the ICHEP 2002, Amsterdam, The Netherlands.}}

\author{Chung Kao\address{Department of Physics and Astronomy,
        University of Oklahoma, \\
        Norman, Oklahoma 73019, USA}%
        \thanks{E-mail address: Kao@physics.ou.edu}}
       
\begin{document}

\begin{abstract}
We discuss the prospects for the discovery of neutral Higgs bosons
($\phi^0 = H^0, h^0, A^0$) in multi-Higgs doublet models 
via their decays into leptons at hadron colliders.  
A major focus will be searching for Higgs bosons 
produced with bottom quarks via Higgs decays into muon pairs
($pp \to b\bar{b}\phi^0 \to b\bar{b}\mu\bar{\mu} +X$)
at the CERN Large Hadron Collider 
within the framework of the minimal supersymmetric model.
\end{abstract}

\maketitle

\noindent{\bf 1. Introduction.} 
In this article we discuss the prospects of discovering neutral Higgs bosons 
via their decays into lepton pairs of the same flavor as well as 
final states with lepton flavor violation.
In particular, we present a summary of our results for the minimal 
supersymmetric Higgs bosons produced with bottom quarks via Higgs decays 
into muon pairs at the CERN Large Hadron Collider (LHC) \cite{hbbmm}.

In a two Higgs doublet model such as the minimal supersymmetric model (MSSM) 
with Yukawa interactions of Model II, 
there are two Higgs doublets $\phi_1$ and $\phi_2$ 
coupling to the $t_3 = -1/2$ and $t_3 = +1/2$ fermions, respectively.  
After spontaneous symmetry breaking, there remain five Higgs bosons:
two charged Higgs bosons $H^{\pm}$,
two neutral CP-even scalars $H^0$ (heavier) and $h^0$ (lighter),
and a neutral CP-odd pseudoscalar $A^0$.
The couplings of $\phi^0 b\bar{b}$ and $\phi\ell\bar{\ell}$ 
($\phi^0 = H^0, h^0, A^0$) 
are enhanced by factors of $1/\cos\beta$.

For large $\tan\beta \equiv v_2/v_1$, 
the $\tau\bar{\tau}$ decay mode \cite{Richter-Was,ATLAS}
is a promising discovery channel for the $A^0$ and the $H^0$ 
at the CERN LHC.
It has also been suggested that Higgs bosons might be observable via
their $b\bar{b}$ decays \cite{hbb1} in a large region
of the ($m_A,\tan\beta$) plane. 
However, simulations for the ATLAS detector concluded that 
detection of the $b\bar{b}$ channel would be difficult \cite{ATLAS,hbb2}.

The LHC discovery potential of the muon pair channel for MSSM Higgs bosons 
was demonstrated by Kao and Stepanov \cite{Nikita,CMS},  
and was later confirmed by the ATLAS collaboration \cite{Richter-Was,ATLAS}. 
In the minimal supergravity model, 
the significance of $pp \to \phi^0 \to \mu\bar{\mu} +X$ is 
greatly improved at large $\tan\beta$ \cite{Vernon} because  
$m_A$ and $m_H$ become small from the evolution of 
the renormalization group equations with large $b\bar{b}\phi^0$ couplings.
However, it is very challenging to search for 
Higgs decays into muon pairs in the Standard Model (SM) \cite{hmm0}.

Furthermore, Higgs decays might lead to final states 
with lepton flavor violation \cite{LFV}.
In the MSSM, radiative corrections via gaugino-sfermion loops can lead to 
lepton flavor violating Higgs decays with a branching ratio approximately 
$B(\phi^0 \to \tau \mu) \sim 4\times 10^{-4}$ 
which would be a $3\sigma$ signal at the LHC. 
In an $E_6-$inspired multi-Higgs model with an abelian flavor symmetry, 
lepton flavor violating effects also arise in Higgs decays 
($\phi^0 \to \tau \mu$) 
that might be observable at the Tevatron and the LHC \cite{LFV}.

In this article, we focus on the prospects of discovering
the MSSM Higgs bosons ($\phi^0 = H^0, h^0, A^0$)
produced with bottom quarks\footnote{
Recently, it was suggested that the production of a Higgs boson 
associated with a bottom quark ($g b -> \phi^0 b$) 
might allow for the suppression of backgrounds \cite{Scott}.} 
via Higgs decays into muon pairs
($pp \to b\bar{b}\phi^0 \to b\bar{b} \mu\bar{\mu} +X$) at the LHC.
The Higgs signal and the SM background are evaluated with realistic cuts.

\medskip

\noindent{\bf 2. Higgs Decays into Muons.}
We calculate the cross section at the LHC for 
$pp \to b\bar{b} \phi^0 +X$ ($\phi^0 = H^0, h^0, A^0$) from two subprocesses 
$gg \to b\bar{b} \phi^0$ and $q\bar{q} \to b\bar{b} \phi^0$.
It turns out that the contribution from quark-antiquark annihilation 
($q\bar{q} \to b\bar{b} \phi^0$) is negligible.

It is a good approximation to evaluate the next to the leading order (NLO) 
cross section of $pp \to \phi^0 b\bar{b} +X$ 
with the leading order (LO) contribution 
multiplied by a K factor\footnote{
The K factor is defined as $K = \sigma_{\rm NLO}/\sigma_{\rm LO}$, 
where $\sigma_{LO}$ is evaluated with LO parton distribution functions and
1-loop evolution of the strong coupling ($\alpha_s$) and
$\sigma_{NLO}$ is evaluated with NLO parton distribution functions and
2-loop evolution of $\alpha_s$.}
of 0.8 \cite{Plumper} and using the pole mass $M_b = 4.7$ GeV. 
The results of Ref. \cite{QCD} are applied to 
compute the NLO rates for $pp \rightarrow \phi^0$.
We take a K factor of 1.5 for the contribution from $gg \to H^0,h^0$ 
for $\tan\beta < 6$ and a K factor of 1.3 for $\tan\beta \ge 6$.
The contribution from $gg \to A^0$ is calculated with the 
LO contribution multiplied by a K factor of 1.7
for $\tan\beta < 6$ and a K factor of 1.3 for $\tan\beta \ge 6$.

The cross section for $pp \to b\bar{b} \phi^0 \to b\bar{b} \mu\bar{\mu} +X$ 
is evaluated with the Higgs production cross section 
$\sigma(pp \to b\bar{b} \phi^0 +X)$ 
multiplied by the branching fraction of the Higgs decay into muon pairs
$B(\phi^0 \to \mu\bar{\mu})$.
When the $b\bar{b}$ mode dominates Higgs decays, 
the branching fraction for $A^0 \to \mu\bar{\mu}$ 
is approximately $3 \times 10^{-4}$ for $m_A = 100$ GeV.
 
\medskip

\noindent{\bf 3. The Discovery Potential at the LHC.} 
The dominant physics backgrounds to the final state of $b\bar{b}\mu\bar{\mu}$ 
come from $gg \to b\bar{b}\mu\bar{\mu}$ and $q\bar{q} \to b\bar{b}\mu\bar{\mu}$
as well as $gg \to b\bar{b}W^+W^-$ and $q\bar{q} \to b\bar{b}W^+W^-$
followed by the decays of $W^\pm \to \mu^\pm \nu_\mu$.
We have applied the same acceptance cuts and efficiencies of $b$-tagging 
and mistagging as those of the ATLAS collaboration \cite{ATLAS}.

We show the invariant mass distribution of muon pairs in Fig. 1
for $pp \to b\bar{b} A^0 +X \to b\bar{b} \mu\bar{\mu} +X$ 
with $\tan\beta = 10$ and 50, 
the SM processes of $pp \to b\bar{b} \mu\bar{\mu} +X$, and
contributions from $pp \to b\bar{b} W^+ W^- +X$.
Also shown are the muon pair invariant mass distributions 
for the inclusive final state of $pp \to A^0 \to \mu\bar{\mu} +X$
and its dominant background from the Drell-Yan process
$pp \to Z,\gamma \to \mu\bar{\mu} +X$.
There are several interesting aspects to note from this figure: 
(a) including two $b$ quarks in the final state greatly improves the 
signal to background ratio with realistic efficiencies of b tagging 
and jet mistagging at the ATLAS and the CMS detectors,
(b) the SM subprocess of $gg \to b\bar{b} \mu\bar{\mu}$ and 
$q\bar{q} \to b\bar{b} \mu\bar{\mu}$ make the major contribution 
to the physics background for $M_{\mu\bar{\mu}} \alt 100$ GeV,
but $gg \to b\bar{b}W^+W^-$ and $q\bar{q} \to b\bar{b}W^+W^-$ 
become the dominant background for higher muon pair invariant mass. 

\begin{figure}[htb]
\centering\leavevmode
\epsfxsize=3in\epsffile{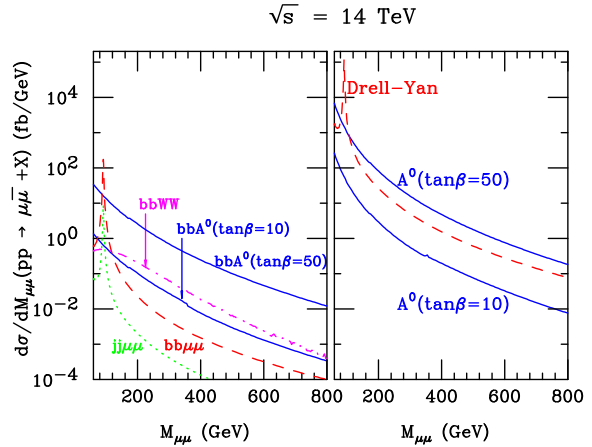}
\caption[]{
The invariant mass distribution for
$pp \to b\bar{b}A^0 +X \to b\bar{b}\mu\bar{\mu} +X$ 
at $\sqrt{s} = 14$ TeV with $M_{\mu\mu} = m_A$.
Also shown are the SM backgrounds from
$pp \to b\bar{b}\mu\bar{\mu} +X$ (dashed), $jj\mu\bar{\mu} +X$ (dotted), 
and $b\bar{b}W^+W^- +X$ (dot-dashed).
The panel on the right shows the invariant mass distribution for the
inclusive final state $pp \to A^0 \to \mu\bar{\mu} +X$ and the Drell-Yan
background (dashed).
\label{fig:sigma}
}\end{figure}

To study the discovery potential of 
$pp \to b\bar{b} \phi^0 +X \to b\bar{b} \mu\bar{\mu} +X$ at the LHC, 
we calculate the background from the SM processes of 
$pp \to b\bar{b} \mu\bar{\mu} +X$
in the mass window of
$m_\phi \pm \Delta M_{\mu\bar{\mu}}$ where 
$\Delta M_{\mu\bar{\mu}} \equiv 
1.64 [ (\Gamma_\phi/2.36)^2 +\sigma_m^2 ]^{1/2}$ \cite{ATLAS},
$\Gamma_\phi$ is the total width of the Higgs boson,  
and $\sigma_m$ is the muon mass resolution.
We take $\sigma_m$ to be $2\%$ of the Higgs boson mass \cite{ATLAS}.
The CMS mass resolution will be better than $2\%$ of $m_\phi$ for 
$m_\phi \alt$ 500 GeV \cite{Nikita,CMS}. 

The 5$\sigma$ discovery contours for the MSSM Higgs bosons at 
$\sqrt{s} =$ 14 TeV 
for an integrated luminosity of $L = 30 \;\; {\rm fb}^{-1}$ and 
$L = 300 \;\; {\rm fb}^{-1}$ are shown in Fig. 2.
We have chosen 
$M_{\rm SUSY} = m_{\tilde{q}} = m_{\tilde{g}} = m_{\tilde{\ell}} 
= \mu = 1$ TeV.

\begin{figure}[htb]
\centering\leavevmode
\epsfxsize=3in\epsffile{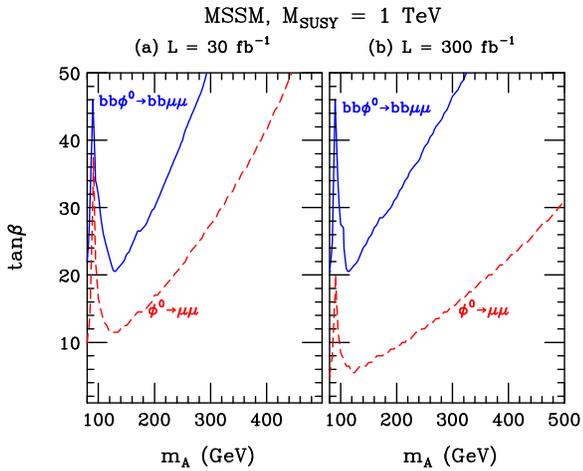}
\caption[]{
The $5\sigma$ contours at the LHC 
for an integrated luminosity ($L$) of 30 fb$^{-1}$ and 300 fb$^{-1}$ 
in the $m_A$ versus $\tan\beta$ plane.  
The signal includes $\phi^0 = A^0$ and $h^0$ for $m_A < 125$ GeV, 
but $\phi^0 = A^0$ and $H^0$ for $m_A \ge 125$ GeV.
The discovery region is above the $5\sigma$ contour.
\label{fig:contour}
}\end{figure}

We have presented results for $M_{\rm SUSY} = 1$ TeV.
If $M_{\rm SUSY}$ is smaller, 
the discovery region of $A^0,H^0 \to \mu\bar{\mu}$ 
will be slightly reduced for $m_A \agt 250$ GeV,
because the the Higgs bosons can decay into SUSY particles 
and the branching fraction of $\phi^0 \to \mu\bar{\mu}$ is suppressed 
\cite{Nikita,Vernon}.

\medskip

\noindent{\bf 4. Conclusions.} 
The muon pair decay mode is a promising channel to discover 
the MSSM Higgs bosons at the LHC. 
The associated final state of $b\bar{b}\phi^0 \to b\bar{b}\mu\bar{\mu}$ 
could discover the $A^0$ and the $H^0$ at the LHC 
with an integrated luminosity of 30 fb$^{-1}$ if $m_A \alt 300$ GeV.
At a higher luminosity of 300 fb$^{-1}$, the discovery region in $m_A$ 
is not expanded much, because the harder $p_T$ cut on $b$ quarks reduces 
the Higgs production cross section.

The inclusive final state of $\phi^0 \to \mu\bar{\mu}$ 
could allow the discovery of the $A^0$ and the $H^0$ at the LHC 
with an integrated luminosity of 30 fb$^{-1}$ if $m_A \alt 450$ GeV.  
At a higher luminosity of 300 fb$^{-1}$, 
the discovery region in $m_A$ is significantly extended to $m_A \alt 650$ GeV 
for $\tan\beta = 50$.

The discovery of both 
$\phi^0 \to \tau\bar{\tau}$ and $\phi^0 \to \mu\bar{\mu}$ 
will allow us to understand the Higgs Yukawa couplings with the leptons.
The discovery of the associated final state of 
$b\bar{b}\phi^0 \to b\bar{b}\mu\bar{\mu}$ 
will provide information about the Yukawa couplings of 
$b\bar{b}\phi^0$ and an opportunity to measure $\tan\beta$. 
Although the discovery region of the $\mu\bar{\mu}$ mode 
is smaller than the $\tau\bar{\tau}$ channel, 
the $\mu\bar{\mu}$ channel allows a precise reconstruction
of the Higgs boson masses. 

\medskip

\noindent{\bf Acknowledgments.}
I am very grateful to Sally Dawson and Duane Dicus for an inspiring 
and enjoyable collaboration. 
This research was supported in part by 
the U.S. Department of Energy under Grant No.~DE-FG03-98ER41066.


\end{document}